\documentclass[prl,showpacs,twocolumn,superscriptaddress]{revtex4}

\usepackage{amsmath,graphicx,latexsym}
\usepackage{bm}

\begin{document}

\title{Electric displacement as the fundamental variable
in electronic-structure calculations}

\author{Massimiliano Stengel}
\affiliation{Materials Department, University of California, Santa Barbara,
             CA 93106-5050, USA}
\author{Nicola A. Spaldin}
\affiliation{Materials Department, University of California, Santa Barbara,
             CA 93106-5050, USA}
\author{David Vanderbilt}
\affiliation{Department of Physics and Astronomy, Rutgers University,
             Piscataway, New Jersey 08854-8019, USA}

\date{\today}

\begin{abstract}
Finite-field calculations in periodic insulators are technically and
conceptually challenging, due to fundamental problems in defining
polarization in extended solids.
While significant progress has been made recently with the
establishment of techniques to fix the electric
field $\bm{\mathcal{E}}$
or the macroscopic polarization $\mathbf{P}$
in first-principles calculations, both methods lack the ease of use and
conceptual clarity of standard zero-field calculations.
Here we develop a new formalism in which the electric displacement
$\mathbf{D}$, rather than $\bm{\mathcal{E}}$ or $\mathbf{P}$, is the
fundamental electrical variable.
Fixing $\mathbf{D}$ has the intuitive interpretation of imposing open-circuit
electrical boundary conditions, which is particularly useful in studying
ferroelectric systems.
Furthermore, the analogy to open-circuit capacitors suggests an appealing
reformulation in terms of free charges and potentials, which
dramatically simplifies the treatment of stresses and strains.
Using PbTiO$_3$ as an example, we show that our technique allows full control
over the electrical variables within the density functional formalism.

\end{abstract}


\maketitle

\def\msm#1{}
\def\dvm#1{}
\def\nsm#1{}


The development of the modern theory of
polarization~\cite{King-Smith/Vanderbilt:1993} has
fueled exciting progress in the theory of the ferroelectric state.
Many properties that could previously be inferred
only at a very qualitative level can now be computed with quantum-mechanical
accuracy within first-principles density-functional theory.
Early \emph{ab-initio} studies focused on bulk ferroelectric crystals,
elucidating the delicate balance between covalency and electrostatics
that gives rise to ferroelectricity.
Over time, these methods were extended to treat the effects of
external parameters such as strains or electric
fields~\cite{Souza/Iniguez/Vanderbilt:2002,Umari/Pasquarello:2002}.
Of particular note is the recent introduction by Di\'eguez
and Vanderbilt~\cite{Dieguez/Vanderbilt:2006} of a method for
performing calculations at fixed macroscopic polarization $\mathbf{P}$.
The ability to compute crystal properties from first principles as
a function of $\mathbf{P}$ provides an an intuitive and appealing link to
Landau-Devonshire and related semiempirical theories in which $\mathbf{P}$
serves as an order parameter.

\begin{figure}
\begin{center}
\includegraphics[width=0.9\columnwidth]{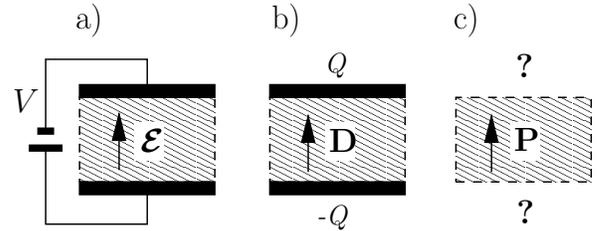}
\end{center}
\caption{ \label{fig1} {\bf Electrical boundary conditions within different
methods.} a) The fixed-$\bm{\mathcal{E}}$ method corresponds to
adopting closed-circuit boundary conditions with a constant applied
bias $V$. b) Constraining $\mathbf{D}$ corresponds to a capacitor in
open-circuit conditions with a fixed value of the free charge $Q$ on the plates.
c) constraining $\mathbf{P}$ does not correspond to a clear experimental
set-up.}
\end{figure}

Despite its obvious appeal, however, the constrained-$\mathbf{P}$
method has found limited practical application to date.
One reason for this is that the procedure to
enforce a constant $\mathbf{P}$ during the electronic self-consistency
cycle is relatively involved; this hampers the study of complex heterostructures
with large supercells, where computational efficiency is crucial.
There are also physical reasons.  In particular,
fixing $\mathbf{P}$ does not correspond to experimentally realizable
electrical boundary conditions (Fig.~\ref{fig1}).
Moreover, in an inhomogeneous heterostructure, the local polarization can
vary from one layer to another, and its average
is therefore best regarded as a derived, not a fundamental, quantity.
In the following we show that considering $\mathbf{D}$ as the fundamental
electrical variable overcomes these physical limitations, and that constraining
$\mathbf{D}$ rather than $\mathbf{P}$ leads to a simpler implementation.

{\it Formalism.}
We consider a periodic insulating crystal defined by three primitive translation
vectors $\mathbf{a}_i$, with $\Omega$ the unit cell volume, and we introduce
the new functional
\begin{equation}
U(\mathbf{D},v) = E_{\rm KS}(v) +
\frac{\Omega}{8\pi} [ \mathbf{D} - 4\pi \mathbf{P}(v) ]^2 \;.
\label{functional}
\end{equation}
$U(\mathbf{D},v)$ depends directly on an external vector parameter 
$\mathbf{D}$, and indirectly on the internal (ionic and electronic) 
coordinates $v$ through the Kohn-Sham energy $E_{\rm KS}$ and the 
Berry-phase polarization $\mathbf{P}$~\cite{King-Smith/Vanderbilt:1993}.
(For the moment we fix the lattice vectors;
strains will be discussed shortly.)
The minimum of $U$ at fixed $\mathbf{D}$ is given by
the stationary point where all the gradients with respect to $v$ vanish,
\begin{equation}
\frac{\partial U}{\partial v} \Big|_{\mathbf{D}} =
\frac{\partial E_{\rm KS}}{\partial v} - \Omega \, (\mathbf{D} - 4\pi \mathbf{P})
\cdot \frac{\partial \mathbf{P}}{\partial v} = 0 \;.
\end{equation}
Comparing with the fixed-$\bm{\mathcal{E}}$ approach of
Ref.~\cite{Souza/Iniguez/Vanderbilt:2002,Umari/Pasquarello:2002} in which the
electric enthalpy $\mathcal{F}$ is given by
\begin{equation}
\mathcal{F}(\bm{\mathcal{E}},v) = E_{\rm KS}(v) - \Omega\, \bm{\mathcal{E}} \cdot \mathbf{P}(v)
\;,
\label{fixede}
\end{equation}
we see that
\begin{equation}
\frac{\partial \mathcal{F}}{\partial v} \Big|_{\bm{\mathcal{E}}} =
\frac{\partial U}{\partial v} \Big|_{\mathbf{D}}
\label{gradeq}
\end{equation}
provided that we set
$\bm{\mathcal{E}} = \mathbf{D} - 4\pi \mathbf{P}$.
We thus discover that $\mathbf{D} = \bm{\mathcal{E}} + 4\pi \mathbf{P}$
is the macroscopic electric displacement field.
The functional in Eq.~(\ref{functional}) takes the form
$U = E_{\rm KS} + (\Omega / 8\pi)\,  \bm{\mathcal{E}}^2$,
which is the correct expression for the internal energy of a
periodic crystal when a uniform external field is present 
(details are given in Supplementary Section 2.4). 
Eq.~(\ref{functional}) thus provides a framework for finding the
minimum of the internal energy $U(\mathbf{D})$ with respect to all internal
degrees of freedom at specified electric displacement $\mathbf{D}$.
This is the essence of our constrained-$\mathbf{D}$ method.

As a consequence of Eq.~(\ref{gradeq}), the method is analogous to a
standard finite-$\mathcal{E}$-field
calculation~\cite{Souza/Iniguez/Vanderbilt:2002,Umari/Pasquarello:2002}.
In particular, the Hellmann-Feynman forces are computed in the same way.
The only difference
is that the value of $\bm{\mathcal{E}}$, instead of being kept constant, is
updated at every iteration until the target value
of $\mathbf{D}$ is obtained at the end of the self-consistency cycle (or ionic
relaxation).
This implies that the implementation and use of the constrained-$\mathbf{D}$
method in an existing finite-$\cal E$-field code is immediate; in our
case it required the modification of \emph{two lines} of code only.

The effect of constraining $\mathbf{D}$, rather than $\bm{\mathcal{E}}$,
essentially corresponds to the imposition of longitudinal, rather than
transverse, electrical boundary conditions.  For example, as we shall
see below, the phonon frequencies obtained from the force-constant
matrix computed at fixed $\mathbf{D}$ are the longitudinal optical
(LO) ones, while the usual approach yields instead the
transverse optical (TO) frequencies.  Furthermore, the longitudinal
electrical boundary conditions are appropriate to the physical
realization of an \emph{open-circuit} capacitor with fixed free
charge on the plates, while the usual approach applies to a closed-circuit
one with a fixed voltage across the plates (Fig.~\ref{fig1}).

{\it Stress tensor.}
This analogy with an open-circuit capacitor suggests an intuitive
strategy for deriving the stress tensor, a quantity that plays a
central role in piezoelectric materials.
In particular, the electrode of an isolated open-circuit capacitor
cannot exchange \emph{free} charge with the environment. This suggests
that the \emph{flux} of the vector field $\mathbf{D}$ through the three
independent facets of the primitive unit cell should remain constant
under an applied strain.
These fluxes are $\mathbf{a}_i\times\mathbf{a}_j\cdot\mathbf{D} =
\Omega\,\mathbf{b}_i\cdot\mathbf{D}$, where the $\mathbf{b}_i$ are
duals ($\mathbf{a}_i\cdot\mathbf{b}_j=\delta_{ij}$) differing by a
factor of $2\pi$ from the conventional reciprocal lattice vectors.
We then rewrite the functionals in terms of the
``internal'' or ``reduced'' variables
$d_i = (\Omega/4\pi)\,\mathbf{b}_i\cdot\mathbf{D}$.  It is also
useful to define the reduced polarization
$p_i=\Omega\,\mathbf{b}_i\cdot\mathbf{P}$
and the ``dual'' reduced electric field
$\bar \varepsilon_i= \mathbf{a}_i\cdot\mathcal{E}$.
Additional details are provided in the Supplementary Section 4.

By Gauss's law, $d_i=-Q_i$, where the $Q_i$ are the
free charges per surface unit cell located
on the cell face normal to $\mathbf{b}_i$.
With these definitions, the internal energy can be rewritten as
\begin{equation}
U(\{d\}) = E_{\rm KS} + \frac{2 \pi}{\Omega} \sum_{ij} \big( d_i - p_i \big)
 \, g_{ij} \, \big( d_j - p_j \big)
\label{qqq}
\end{equation}
where we have introduced the metric tensor
$g_{ij}=\mathbf{a}_i\cdot\mathbf{a}_j$.
We then define the fixed-$\{d\}$ stress tensor as
\begin{equation}
\sigma_{\mu\nu} = \frac{1}{\Omega} \left( \frac{dU}{d\eta_{\mu\nu}}
\right)_{\{d\}}
\label{stress-a}
\end{equation}
where $\eta_{\mu\nu}$ is the strain tensor.
By a Hellmann-Feynman argument (see Supplementary Section 4.3)
the total derivative in Eq.~(\ref{stress-a}) can be replaced
by a partial derivative. Using $dg_{ij}/d\eta_{\mu\nu}=2a_{iu}a_{jv}$,
we find
\begin{equation}
\sigma_{\mu\nu} = \sigma_{\mu\nu}^{\rm KS} + \sigma_{\mu\nu}^{\rm Max}
  +  \sigma_{\mu\nu}^{\rm aug} \, ,
\end{equation}
where $\sigma_{\mu\nu}^{\rm KS}$ is the standard zero-field expression,
\begin{equation}
\sigma_{\mu\nu}^{\rm Max} =
\frac{2 \mathcal{E}_\mu \mathcal{E}_\nu
    - \delta_{\mu\nu}\mathcal{E}^2}{8\pi}
\end{equation}
is the Maxwell stress tensor (which originates from the derivative
acting on $g_{ij}$ and $\Omega^{-1}$), and
\msm{Fixed the formula for the augmented stress, the prefactor was wrong.}
\begin{equation}
\sigma_{\mu\nu}^{\rm aug} =
-\frac{1}{\Omega}\sum_{i}
\, \bar \varepsilon_i \,\frac{\partial p_i}{\partial\eta_{\mu\nu}} \,
\end{equation}
is the ``augmented'' part.
If the internal variables $v$ are chosen as reduced atomic
coordinates and plane-wave coefficients in a norm-conserving
pseudopotential context, neither the ionic nor the Berry-phase
component of $p_i$ has any explicit dependence on strain,
and $\sigma_{\mu\nu}^{\rm aug}$ vanishes.  The name thus refers
to the fact that $\sigma_{\mu\nu}^{\rm aug}$ is nonzero only in
ultrasoft pseudopotential \cite{Vanderbilt:1990} and
projector augmented-wave \cite{Bloechl:1994} contexts.

We note that, as a consequence of fixing the reduced variables
$d_i$ rather than the Cartesian $\mathbf{D}$, the
\emph{proper} treatment of piezoelectric
effects~\cite{Vanderbilt:2000,Wu/Vanderbilt/Hamann:2005}
is automatically enforced.
This formal simplification allows for an enhanced flexibility in
the simultaneous treatment of electric fields and strains.
For example, it is possible to introduce a rigorous 
constant-pressure enthalpy by simply defining
\begin{equation}
U^\pi(d) = \min_\eta [U(d,\eta) - \pi \Omega] \, ,
\label{eqpress}
\end{equation}
where $\pi$ is the external pressure and $\Omega$ is the cell volume.
We will demonstrate the use of this strategy in the application to
PbTiO$_3$.

{\it Legendre transformation.}
The transformation from variables $\mathbf{D}$ to variables
$\bm{\mathcal{E}}$ can be regarded as part of a Legendre
transformation.  We spell out this connection here, working instead
with reduced variables $(d_1,d_2,d_3)$ and
$(\bar \varepsilon_1,\bar \varepsilon_2,\bar \varepsilon_3)$.
First, we note that
\begin{equation}
\frac{d U}{d d_i} = \frac{\partial U}{\partial d_i} = \bar \varepsilon_i \, .
\label{Uderiv}
\end{equation}
Recall that $\bar \varepsilon_i = \mathbf{a}_i \cdot \bm{\mathcal{E}}$,
so that $-\bar \varepsilon_i$ is just the potential step $V_i$ encountered
while moving along lattice vector $\mathbf{a}_i$, while $Q_i$ is just
the free charge on cell face $i$.
Thus, when the system undergoes a small change at fixed
$(\bar \varepsilon_1,\bar \varepsilon_2,\bar \varepsilon_3)$, the work done
by the battery is $-\sum_i V_i \, dQ_i = -\sum_i \bar \varepsilon_i \, dd_i$.
We therefore define
\begin{equation}
\tilde{\mathcal{F}}(\bar \varepsilon_1,\bar \varepsilon_2,\bar \varepsilon_3) =
\min_{d_1,d_2,d_3} \big[ U(d_1,d_2,d_3) - \sum_i \bar \varepsilon_i \, d_i \big],
\end{equation}
where the potentials $\bar \varepsilon_i$ have become the new independent
variables and $d_i$ are now implicit in the minimum condition.
The energy functionals $U(\{d_i\})$ and
$\tilde{\mathcal{F}}(\{\bar \varepsilon_i\})$ thus form a
Legendre-transformation pair.

All the gradients with respect to the internal and strain degrees of freedom
are preserved by the Legendre transformation and need not be rederived
for $\tilde{\mathcal{F}}$.
The physical electrical boundary conditions, however, have changed back to
the \emph{closed-circuit} case.
It is therefore natural to expect the functional $\tilde{\mathcal{F}}$ to
be closely related to the fixed-$\bm{\mathcal{E}}$ enthalpy $\mathcal{F}$
of Eq.~(\ref{fixede}).
Indeed, it is straightforward to show that
\begin{equation}
\tilde{\mathcal{F}} = U - \frac{\Omega}{4\pi} \mathcal{E} \cdot \mathbf{D}
                    = \mathcal{F} - \frac{\Omega}{8\pi} \mathcal{E}^2 \;.
\end{equation}
At fixed strain and $\bar \varepsilon_i$, the $\Omega\mathcal{E}^2/8\pi$
term is constant, and thus does not contribute to the gradients
with respect to the internal variables, consistent with Eq.~(\ref{gradeq}).
However, the stress derived from $\mathcal{F}$ differs from
the one derived from $\tilde{\mathcal{F}}$ by the Maxwell
term $\sigma_{\mu \nu}^{\rm Max}$, which is absent in
$\mathcal{F}$ (details of the derivation are
provided in Supplementary Section 4.3). 
Although the Maxwell stress is tiny on the scale of typical
first-principles calculations (e.g. $10^8$ V/m produces a pressure of
$44.3$ KPa), for reasons of formal consistency we
encourage the use of $\tilde{\mathcal{F}}$ in place of $\mathcal{F}$
in future works.

{\it Partial Legendre transformations.}
It is also possible to define hybrid thermodynamic functionals via
partial Legendre transformations that act only on one or two
of the three electrical degrees of freedom.  Of most interest
is the case of two fixed $V$ and one fixed $Q$, i.e., functions
of variables $(\bar \varepsilon_1,\bar \varepsilon_2,d_3)$.  The
special direction is denoted by unit vector $\hat{\mathbf{q}}$ which is
along direction $\mathbf{b}_3$.
When $\bar \varepsilon_1=\bar \varepsilon_2=0$, this
applies to two common experimental situations: the case of an
insulating film sandwiched in the $\hat{\mathbf{q}}$ direction
between parallel electrodes in open-circuit boundary conditions, and the
case of a long-wavelength LO phonon of wavevector $\mathbf{q}$ where the
$\mathbf{q}\rightarrow0$ limit is taken along direction $\hat{\mathbf{q}}$.

This latter case of LO phonons exemplifies the physical
interpretation of our method and its usefulness.  While the {\it
gradients} of $U$ and its partially Legendre-transformed partner
are identical, the {\it force constant matrices}, which are second
derivatives, are not.  Indeed, the force-constant matrices are found
to differ by
\begin{equation}
\Delta K_{I\alpha,J\beta} =
\frac{4\pi}{\Omega} \, \frac{ (\mathbf{Z}_I\cdot\hat{\mathbf{q}})_\alpha
(\mathbf{Z}_J\cdot\hat{\mathbf{q}})_\beta}
{\hat{\mathbf{q}}\cdot\mathbf{\epsilon}_\infty \cdot \hat{\mathbf{q}}} \;,
\label{DKna}
\end{equation}
where $I\alpha$ labels the atom $I$ and its displacement direction
$\alpha$, $\mathbf{Z}_{I\alpha}$ is the corresponding dynamical
charge, and $\mathbf{\epsilon}_\infty$ is the purely electronic
dielectric tensor.  This is 
readily identified as the non-analytic contribution to the
LO-TO splitting of a phonon of small wavevector $\mathbf{q}$ in the
theory of lattice dynamics~\cite{Baroni/deGironcoli/DalCorso:2001}.
This demonstrates that the lattice dynamical-properties of a
given insulating crystal within our fixed-$D$ method are fully 
consistent with what should be expected from a change in the 
electrical boundary conditions from transverse to longitudinal.

{\it Dielectric tensor and linear response.} This scheme lends itself
naturally to the perturbative linear-response
analysis of the second derivatives of the internal energy as
described in Ref.~\cite{Wu/Vanderbilt/Hamann:2005}, with
two important differences.
First, in our scheme the derivatives at constant $\mathbf{D}$ become
the elementary tensors, while the derivatives at constant $\bm{\mathcal{E}}$
are ``second-level'' quantities; this is an advantage, since using $\mathbf{D}$ as
independent variable is very convenient in ferroelectric systems.
Second, the use of the reduced field variables $d_i$ and $\bar{\varepsilon}_i$
in place of the macroscopic vector fields $\mathbf{P}$ and $\bm{\mathcal{E}}$
makes the discussion of strains under an applied field much more rigorous
and intuitive.

As an example of the relationship between constrained-$\bar \varepsilon$ 
and constrained-$d$ tensors it is useful to introduce the inverse capacitance, 
$\gamma=C^{-1}$, in matrix form as
\begin{equation}
\gamma_{ij} = \frac{d^2 U}{d d_i d d_j} \;.
\label{capa}
\end{equation}
Incidentally, while this expression is fully general and
well-defined in the non-linear regime, for the special case of
a linear medium we can write
\begin{equation}
U= U_0 + \frac{1}{2} \sum_{ij} \gamma_{ij} \, Q_i \, Q_j,
\end{equation}
which generalizes the textbook formula
$U= Q^2 / 2C$ to the case of three mutually coupled capacitors.
It can be shown that the same information can be obtained within
the constrained-$\bar \varepsilon$ approach by means of the relationship
\begin{equation}
(\gamma^{-1})_{ij} = \frac{d^2 \tilde{\mathcal{F}}}
   {d \bar \varepsilon_i d \bar \varepsilon_j} \;.
\end{equation}
The matrix $\gamma_{ij}$ can be thought of a ``reduced'' representation of the
macroscopic dielectric tensor,
\begin{equation}
(\epsilon^{-1})_{\alpha \beta} = \frac{\Omega}{4\pi}
\sum_{i,j} \gamma_{ij} \, b_{i,\alpha} \,  b_{j,\beta} \;,
\end{equation}
or equivalently
\begin{equation}
\epsilon_{\alpha \beta} = \frac{4\pi}{\Omega}
\sum_{i,j} (\gamma^{-1})_{ij} \, a_{i,\alpha} \, a_{j,\beta} \;.
\end{equation}

We will consider, in addition to the \emph{total} static capacitance
above, the closely related frozen-strain $\gamma^\eta_{ij}$ and
frozen-ion $\gamma^\infty_{ij}$ tensors.
The remainder of the response functions discussed in
Ref.~\cite{Wu/Vanderbilt/Hamann:2005} can be similarly
defined in terms of the second derivatives of $U(\{d\},u,\eta)$.

{\it Applications.}
In the following we illustrate our method by computing the
electrical equation of state of a prototypical ferroelectric
material, PbTiO$_3$.  Our calculations are performed within the
local-density approximation~\cite{Perdew/Wang:1992} (LDA) of
density-functional theory using norm-conserving~\cite{troullier}
pseudopotentials and a planewave cutoff of 150~Ry.  The tetragonal
unit cell contains one formula unit, and a $6 \times 6
\times 6$ Monkhorst and Pack~\cite{Monkhorst/Pack:1976}
grid is used to sample the Brillouin zone.  The finite
electric field is applied through a Wannier-based real-space
technique~\cite{Stengel/Spaldin:2007},
which converges quickly as a function of $k$-point mesh
resolution~\cite{Stengel/Spaldin:2005}; indeed, tests made with
finer meshes showed no differences within numerical accuracy.
We obtain an equilibrium lattice constant of $a$=3.879 \AA\ for cubic
paraelectric PbTiO$_3$, in line with values previously reported in the
literature.
Due to the tetragonal symmetry, the state of the system is fully determined
by six parameters: the electric displacement $d$, the cell parameters $a$
and $c$, and three internal coordinates describing relative
displacements along $z$.

\begin{figure}
\begin{center}
\includegraphics[width=0.9\columnwidth]{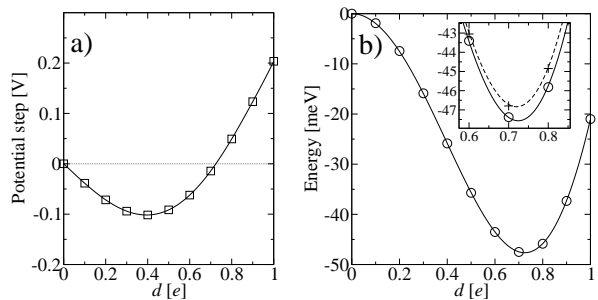}
\end{center}
\caption{ \label{fig2}
   {\bf Potential step and internal energy as a function of $d$.}
   Symbols, calculated using constrained-$\mathbf{D}$ method:
   (a) reduced electric field $\bar{\varepsilon}$ (squares);
   (b) internal energy $U$ (circles).  Solid curves: (a) numerical
   cubic spline fit to the symbols; (b) numerical integral of the
   spline fit in (a).  Inset: enlargement near
   the minimum, also showing magnitude of the error made if
   Pulay stresses are neglected (dashed curve).}
\end{figure}

Starting from the relaxed cubic structure in zero field, we calculate
the equilibrium state for ten evenly spaced values of the reduced
displacement $d$, ranging from $d$=0.1\,$e$ to $d$=1.0\,$e$ (where $-e$
is the electron charge), and relaxing all the structural variables at
each $d$ value. We set a stringent accuracy threshold of $10^{-5}$
Ha/bohr for atomic forces and $10^{-7}$ Ha/bohr$^3$ for stresses.
First we check the internal consistency of the formalism by verifying
that our calculated potential drop $\bar{\varepsilon}$
coincides with the numerical derivative of $U$ with
respect to $d$ as expected from Eq.~(\ref{Uderiv}).
The comparison is shown in Fig.~\ref{fig2}, where the discrepancies, of order
10$^{-6}$\,Ha, are not even visible.  This confirms the internal consistency
of the formalism and the high numerical accuracy of the calculations.
The minimum in Fig.~\ref{fig2} (b) [which coincides with the zero-crossing 
in (a)] at $d=0.725\,e$ corresponds to a spontaneous polarization of
$P_{\rm s}=0.78\,$C/m$^2$.

We note that this comparison is sensitive to the Pulay stress,
even in the present case where our conservative 
choice of the plane-wave cutoff makes this error as small as 
$\pi_P$ = 82 MPa. 
Neglecting such error corresponds to applying a spurious hydrostatic pressure of
$-\pi_P$, which leads to a discrepancy between the integrated potential
(dashed curve in the inset of Fig.~\ref{fig2}) and the calculated internal
energy values.
The agreement can be restored by plotting, instead of $U$ (circles), the 
correct functional, Eq.~(\ref{eqpress}), for constant-pressure conditions
(plus symbols).
As such, this comparison constitutes a stringent test that
all numerical issues have been properly accounted for, particularly in systems like
PbTiO$_3$ that are characterized by a strong piezoelectric response.

\begin{figure}
\begin{center}
\includegraphics[width=0.9\columnwidth]{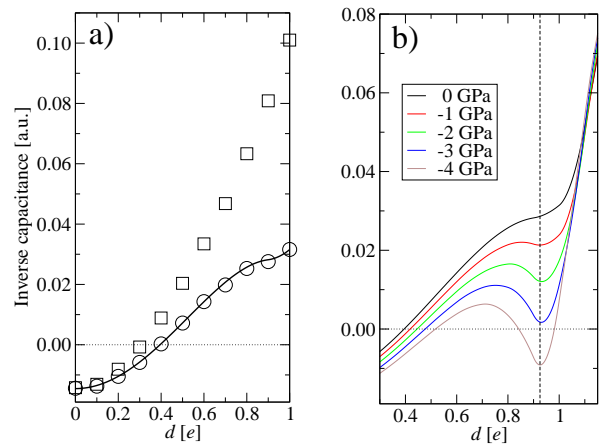}
\end{center}
\caption{ \label{fig3} 
{\bf Dielectric properties.}
(a) Calculated inverse capacitance in the
free-stress ($\gamma$, circles) and fixed strain
($\gamma^{\eta}$, squares) limits. The points were obtained by
extracting the symmetric $6 \times 6$ elementary response tensors
by finite differences (steps of $\pm$0.001 were taken for
each parameter) for each value of $d$; the continuous curve is the
result of numerical differentiation of the splined potential. 
(b) Evolution of $\gamma$ for increasingly larger negative pressures.}
\end{figure}

Having verified the accuracy and consistency of our method, we now demonstrate
its utility by analyzing the second derivative of the internal energy (or
equivalently the first derivative of the potential) as a function of $d$,
which corresponds to the inverse capacitance $\gamma$.
The symbols in Fig.~\ref{fig3}(a) show the linear-response values of
both $\gamma^{\eta}$ and $\gamma$, which are identical in the 
non-piezoelectric cubic limit. The numerical derivative of the 
splined potential of Fig.~\ref{fig2}(a) accurately matches $\gamma$,
again confirming the high numerical quality of our
calculations.
Fig.~\ref{fig3}(a) shows that the inverse capacitance is \emph{negative} for 
$0<d<0.395$ [the zero-crossing point corresponds to the
inflection point of the $U(d)$ curve of Fig.~\ref{fig2}(b), and to the minimum 
of $\bar \varepsilon(d)$ of Fig.~\ref{fig2}(a)].
This is indicative of the fact that cubic PbTiO$_3$ is characterized by a
ferroelectric instability, which means that the $U(d)$ curve has a negative
curvature around the saddle point $d=0$.
We suggest, therefore,  that the constrained-$\mathbf{D}$ inverse capacitance at $d=0$,
while not accessible experimentally (since it corresponds to an unstable
configuration of the crystal), is a useful indicator of the \emph{strength} 
of the ferroelectric instability.
As such, it can play an important role in determining the \emph{critical thickness} for
ferroelectricity in thin perovskite films;
in particular, a material with lower $\gamma$ should
be ferroelectric down to smaller thicknesses, provided that the 
depolarizing effects due to the ferroelectric/electrode interface
are equally important~\cite{Junquera/Ghosez:2003}. 
Note that in our terminology one ferroelectric can be both
\emph{stronger} and {\it less polar} than another
if it has a more negative $\gamma$ but a smaller $|P_s|$.

\begin{figure}
\begin{center}
\includegraphics[width=0.9\columnwidth]{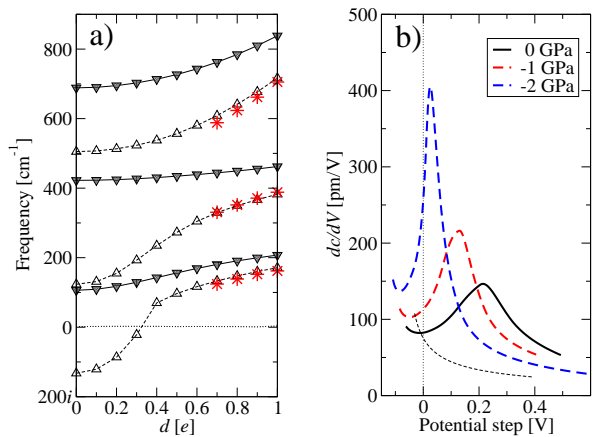}
\end{center}
\caption{{\bf Lattice-dynamical and piezoelectric properties.}
 \label{fig4}(a) Longitudinal (filled symbols,
continuous curves) and transverse (open symbols, dashed curves)
optical modes of $\Gamma_{15}$ symmetry as a function of $d$ at zero
pressure. Selected TO frequencies for $\pi$=-4 GPa are shown as 
red star symbols. (b) Calculated piezoelectric coefficient as a function of
external pressure $\pi$ and bias $V$ (thick curves). 
Results of the Landau-Devonshire model of Refs.~\onlinecite{haun_pbtio3} 
and~\onlinecite{chen_piezo:2003} are shown as a thin dashed 
curve for comparison.}
\end{figure}

\emph{Piezoelectricity.} Interestingly, the free-stress $\gamma(D)$ curve 
in Fig.~\ref{fig3}(a) shows a peculiar plateau for $0.8<d<1.0$,
indicative of a ``softening'' 
of the response of the crystal to the applied electric field.
This behavior is surprising, as an electric field of increasing
strength would rather be expected to drive a perovskite 
crystal further into the anharmonic regime, with a consequent 
progressive hardening of the overall dielectric 
response~\cite{chen_piezo:2003}.
To analyze this effect, we start by observing that the  
evolution of the fixed-strain $\gamma^{\eta}$ dielectric 
response as a function of $d$ is essentially featureless.
This indicates that optical phonons alone cannot be responsible for the
effect, and volume (and/or cell-shape) relaxations are crucially 
involved.
To investigate the coupling between the volume and the 
dielectric response, we repeated our calculations within
a negative hydrostatic pressure by using the mixed fixed-$D$, 
fixed-$\pi$  enthalpy defined in Eq.~(\ref{eqpress}).
The results for the free-stress $\gamma(D)$, plotted in Fig.~\ref{fig3} (b), 
show a dramatic influence of the external pressure on the dielectric response 
of the system.
In particular, the plateau in $\gamma(D)$ becomes an increasingly
deeper local minimum, which crosses the $\gamma=0$ axis for $\pi$ between 
$-$3 and $-$4\,GPa; the local minimum is approximately centered in $d=0.925$
for all values of $\pi$.

A negative $\gamma$ indicates a structural instability, and structural
instabilities in ferroelectric systems are usually understood in terms
of ``soft'' phonon modes. In order to see whether this picture applies 
here, we plotted in Fig.~\ref{fig4}(a) the zone-center polar mode frequencies  
as a function of $d$.
At zero pressure, the curves show no notable irregularity, consistent
with the smooth evolution of the fixed-strain response $\gamma^\eta$.
Remarkably, the external pressure has a negligible influence on 
such frequencies, which remain practically unchanged for the strained 
crystal at the same value of $d$, confirming our hypothesis that the
effect is essentially of piezoelectric nature.

Pursuing this idea, we combined the values of the 
potential drop $V(d)$ with the equilibrium values of out-of-plane 
lattice parameter, and calculated the free-stress piezoelectric
coefficient by numerical differentiation as
\begin{equation}
h = -\frac{dc}{dV} \,.
\end{equation}
The results, plotted in Fig.~\ref{fig4}(b), show for zero pressure a clear peak
at $V \sim 0.2 V$, corresponding to a value of the internal field of about 
450 MV/m. We identify this peak with the plateau in the inverse capacitance 
curve of Fig.~\ref{fig3}(a).
For incresingly large negative pressures, the piezoelectric peak becomes
sharper and shifts to smaller values of the potential; for $\pi<-3$\,GPa
(not shown) the piezoelectric coefficient diverges, corresponding to the
crossover to the unstable region in Fig.~\ref{fig3} (b).

These results shed light on the recent experimental 
measurements of Ref.~\onlinecite{grigoriev}, where a remarkable
anomaly in the piezoelectric response of PZT films in high fields
($\mathcal{E}=200-300$ MV/m) was detected.
Such an anomaly was rationalized in terms of a transition to a supertetragonal
state, which previous first-principles calculations~\cite{tinte} predicted 
to be stable in PbTiO$_3$ under an applied negative hydrostatic pressure. 
However, the question remained whether the electric field might actually 
produce such a transition within the experimental range of applied
fields. Our results demonstrate that the piezoelectric coupling is
indeed capable of driving such a transition, at least under isotropic
free-stress boundary conditions. 
(In future work, we plan to extend
these investigations by considering epitaxial strain clamping effects.)
We note that our calculated value of the piezoelectric coefficient of 
PbTiO$_3$ at zero field and pressure ($h=82.5$ pm/V) is in excellent 
agreement with previous Landau-Devonshire theories~\cite{haun_pbtio3,
chen_piezo:2003}, but the evolution of $h$ for nonzero values of
the applied potential substantially differs [see Fig.~\ref{fig4} (b)].
For small values of the electric field, in particular, our \emph{ab-initio}
results indicate that $h$ remains roughly constant. Then $h$ increases
significantly at higher fields, up to a value $\mathcal{E}\sim$ 450\,MV/m,
where it starts decreasing again. A monotonic decrease was predicted 
instead by the model of Ref.~\onlinecite{chen_piezo:2003}.
This result, therefore, has important implications for the tunability 
of the piezoresponse of lead titanate crystals.

{\it Summary and outlook.} In conclusion, we have presented a formalism that
provides full control over the electrical degrees of freedom in
a periodic first-principles electronic-structure calculation.
We have in mind several immediate applications for our
method. First and foremost, fixing $D$ in ferroelectric capacitors
by using the methods of Ref.~\onlinecite{Stengel/Spaldin:2007} will allow
for a detailed analysis of the microscopic mechanisms determining
the depolarizing field, both in the linear and anharmonic regime,
an issue which is central to the development of efficient ferroelectric 
devices. 
Second, imposing constant-$D$ electrical boundary conditions has 
the virtue of making the force-constant matrix of layered heterostructures 
\emph{short-ranged} in real space. This allows one to accurately model
the polarization and response of complex superlattices, capacitors and 
interfaces in terms of the electrical properties of the elementary
building blocks; a demonstration of this strategy was recently reported in
Ref.~\onlinecite{xifan_2008}.
Third, complex couplings between different order parameters can
now be treated with unprecedented flexibility,
opening new avenues in the
theoretical study of magnetoelectric multiferroics and improper 
ferroelectrics.

{\it Acknowledgements.} This work was supported by the Department of 
Energy SciDac program on Quantum Simulations of Materials and 
Nanostructures, grant number DE-FC02-06ER25794 (M.S. and N.A.S.), 
and by ONR grant N00014-05-1-0054 (D.V.).


\end{document}